\documentclass[]{article}

\usepackage{graphicx}
\begin{document}

\bibliographystyle{plain}

\title{An ant-based algorithm for annular sorting}

\author{Oliver Don and Martyn Amos \\ Department of Computer Science, University of Exeter, \\ Exeter EX4 4QF, United Kingdom \\ \\ Email: M.R.Amos@exeter.ac.uk}

\maketitle
\begin{abstract}
In this paper we describe a minimal model for annular sorting by {\it Leptothorax} ants. Simulation results are consistent with the structures observed in actual ant colonies.
\end{abstract}

\section{Introduction}

The ability of social insects to collectively solve problems has been well-studied and documented \cite{camazine01}. The behaviour of foraging ants, for example, has been abstracted to provide algorithmic solutions that are robust, distributed and flexible \cite{dorigo99,dorigo04}.  The particular behaviour that we will focus on is the clustering or sorting of ant corpses or larvae \cite{den91}. Abstract models of these behaviours have been successfully applied to, amongst other problems, numerical data analysis, data mining and graph partitioning \cite{handl03antbased}.
In this paper, we focus on the task of {\it brood sorting}. This behaviour, when observed in {\it Leptothorax unifasciatus} \cite{franks02}, leads to the formation of a single cluster of offspring made up of concentric rings of brood items, with the youngest items (eggs and micro-larvae) being tightly packed at the centre, and successively larger larvae arranged in increasingly wider-spaced bands moving out from the centre of the cluster.

In this paper we investigate an intriguing hypothesis by Franks et al. concerning the biological mechanisms underlying annular sorting. In their article, the authors state that ``The mechanism that the ants use to re-create these brood patterns when they move to a new nest is not fully known. Part of the mechanism may involve conditional probabilities of picking up and putting down each item which depend on each item's neighbours \ldots The mechanisms that set the distance to an item's neighbour are unknown. They may be pheromones that the brood produce and which tend to diffuse over rather predicable distances \ldots" \cite{franks02}

We have constructed a corresponding minimal model, using only stochastic ant behaviour and  pheremone concentrations, that appears to account for the emergence of annular clusters of objects in simulated {\it Leptothorax} colonies. In Section 2 we present the background to the problem, before describing our model in
Section 3. We conclude with a summary of our results and discussion of their implications.

\section{Annular sorting}

\begin{figure}
\begin{center}
\includegraphics[scale=0.75]{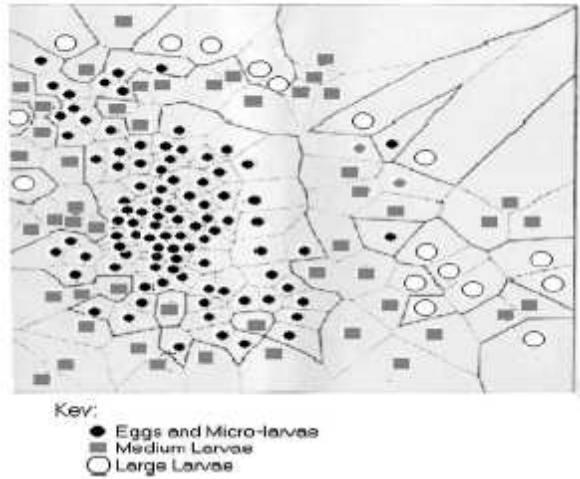}
\caption{Annular sorting in a real {\it Leptothorax} colony (taken from \cite{franks02}, with permission).}
\label{fig:franks}
\end{center}
\end{figure}

Franks {\it et al.} \cite{franks02} carried out an observational biological study of the brood 
sorting behaviour of \textit{Leptothorax} ants. The nesting behaviour of this species made them 
particularly suitable for study as they nest in single clusters in flat rock crevices, a situation that is easy
to replicate and monitor in a laboratory. Photographs were taken of the ants' brood cluster and individual items in the  cluster were classified, before a tessellation was applied to  determine density and distance from the cluster
centroid. This study showed that \textit{Leptothorax} ants placed  smaller brood items at the centre of the cluster with a greater density, forming an annular cluster. This arrangement reasserted itself when the ants 
were forced to migrate to a new nesting site, and proved to be an ongoing  process. This structure is illustrated in Figure ~\ref{fig:franks}, where three different types of brood item are arranged in roughly-sorted concentric rings.

Wilson \textit{et al.} have recently proposed the first model of ``ant-like annular sorting'' to 
simulate the behaviour of \textit{Leptothorax} ants using minimalist robot and computer 
simulations \cite{wilson04}. 
Three models for annular sorting were presented: ``Object clustering using objects of different size'', ``Extended differential pullback'' and `` Combined leaky integrator. The first was run 
exclusively as a computer simulation, since modifying robots to allow them to 
move objects of different sizes proved to be too complex. Despite this, the 
computer simulation modelled physical robot  behaviour faithfully, preserving the limitations of movement 
inherent in simple robots, and even going so far as to build in a 1{\%} sensor error that 
matched the rate seen in the machines. 

The first model explored the theory that annular sorting occurs solely due 
to the different sizes of the objects involved, Wilson \textit{et al.} compared this 
to the manner in which muesli settles in transit, with smaller clusters 
falling to the bottom, leaving the larger ones on top. The simulation 
modelled agents who picked up the first object they encountered while 
unladen and deposited it at the next object they encountered. The results 
displayed a slight increase in the quality of clusters when there was a 
greater number of different object sizes; however, clusters tended to form at the edge of the 
area and were often ``inside out", i.e. with larger objects at the centre 
surrounded by smaller objects. This lead Wilson \textit{et al.} to conclude that 
merely using objects of different sizes did ``not create a sufficient muesli 
effect''.

The second and third models attempted to recreate ant brood clustering 
behaviour by assigning more complex behaviour to the ants. Wilson \textit{et al.}
hypothesised that ants were able to recognise inherent differences in larval 
growth, and created annular clusters by depositing different objects at 
different differences away from each other depending on size. Due to the 
limitations of the robots, this was implemented by having the agents reverse 
a distance depending on the kind of object carried before depositing it. 
Initial results with this approach were not good. As a result Wilson \textit{et al.} 
proposed a third model, calling it the ``Leaky integrator''. This allowed 
the agents to have an adaptive amount of ``pullback" that varied according to 
how many objects of the same kind had been encountered in the last $n$
seconds. Initial tests with this system produced poor results but results 
improved when a genetic algorithm was used to select parameter values.

\section{Our proposed model}

We now propose an alternative model to explain the phenomenon of annular clustering.
Ants are represented as ``agents'', and brood items of different sizes are 
represented by ``objects''. Agents and objects are spatially-distributed randomly on a 
two dimensional ``board" of fixed dimensions.

Each object has a {\it placement score}; agents move randomly across the board, and  
when they collide with an object they calculate its placement score. This score is 
then used to probabilistically determine whether the agent should {\it pick up} 
the object and become laden. Laden agents carry objects around the board, and at
every timestep they evaluate what placement score the carried object {\it would have} if it 
were to be deposited at the current point. This score is then used to probabilistically determine 
whether the object should be {\it deposited}.

Each category of object to be clustered posses three attributes: {\it size, 
minimum perimeter} and {\it maximum perimeter}. Size is important, as objects may not 
overlap, the other two attributes are both functions of the object's size. 
Minimum and maximum perimeter are used in the calculation of the object's 
placement score: When evaluating the placement score of an object, the agent 
counts how many nearby objects fall within the minimum perimeter (these count towards
a weighted penalty), and how many  fall within the maximum perimeter (these count towards a weighted bonus). 

\begin{figure}
\begin{center}
\includegraphics[scale=0.4]{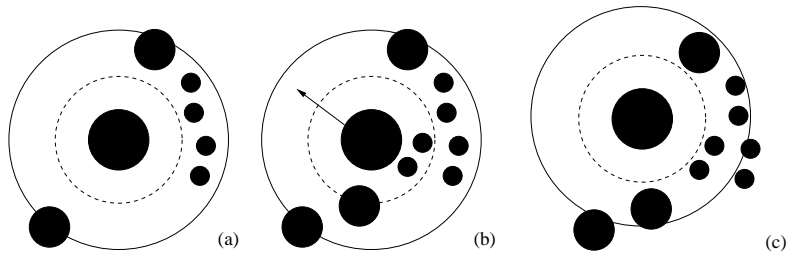}
\caption{Depiction of how large objects are forced out of clusters. (a) Good configuration. (b) Placement score of central object drops as a result of three new objects being placed within its minimum perimeter. (c) Object is replaced with a higher placement score.}
\label{fig:repulsion}
\end{center}
\end{figure}

It is important to note that the maximum and minimum perimeters are only 
calculated for the object whose placement score is currently being 
evaluated. Agents take no account of whether a placement results in a good 
or bad score for neighbouring objects, thus small objects with smaller 
minimum perimeters will frequently be placed quite close to larger objects, 
resulting in a penalty score for those larger objects. As a result groups of 
smaller objects will tend to ``force out" larger objects. It could be argued 
that this process implements a stronger version of the size-based muesli 
analogy proposed by Wilson \textit{et al.} \cite{wilson04}
As stated earlier, the minimum and maximum perimeter values are a function of 
an object's size (we describe the exact function shortly). The minimum perimeter of an object acts as a
 ``repulsive" force, whereas the maximum perimeter acts as an ``attractive" force. These coupled
forces ensure that objects maintain a proportional ``exclusion zone",
around themselves, whilst ensuring the coherence of a single cluster around some centroid
(``centre of gravity"). We now illustrate this with an example, depicted in Figure ~\ref{fig:repulsion}.

This example shows how larger objects are ``forced" out of clusters by their proximity to other objects. The large object in Figure ~\ref{fig:repulsion} has its minimum perimeter depicted as a dotted line, and its maximum perimeter as a solid line. In Figure ~\ref{fig:repulsion}(a), the object has a high-scoring placement score, as it has no objects within its minimum perimeter (which would attract a penalty), and several objects in its maximum perimeter (attracting a bonus score). However, if several objects are later deposited in the central object's immediate vicinity (Figure ~\ref{fig:repulsion}(b), its placement score changes for the worse, as these objects contribute a significant penalty. As a result, the next time this object is encountered by an agent it will be carried until it once again has a beneficial placement score (Figure ~\ref{fig:repulsion}(c), where it will be deposited. As we can see, this calculation of placement scores has the effect of moving larger objects away from clusters of smaller objects, whilst maintaining relative proximity.

\begin{figure}
\begin{center}
\includegraphics[scale=0.3]{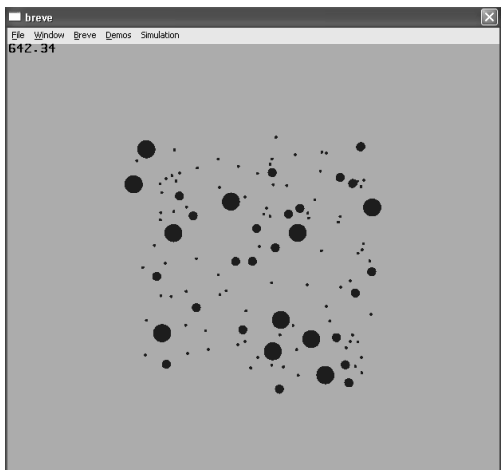}
\caption{Poor clustering caused by zero delay}
\label{fig:poor}
\end{center}
\end{figure}

Before embarking on a full implementation of our algorithm, we carried out test runs using a prototype system.
This highlighted two, previously unforseen, effects that we deal with in the full algorithm.
Early investigations showed that ants had a tendency to construct good annular clusters, but would then deconstruct them at the outer edge, as there was no termination criterion defined for the algorithm. We solved this problem by introducing the notion of ``energy"; each ant starts with a fixed amount of energy, represented as an integer value, which is decremented every time the ant picks up an object. Once the agent's energy level reaches zero is is removed from the system, and the simulation terminates when there are no more agents left. This prevents the simulation from ending when objects are still being carried, but also has the beneficial effect of smoothly and gradually reducing the overall activity within the system as the simulation approaches termination. The biological validity of this approach is unclear, but it appears to make the clustering of outer objects much more realistic.

Another problem that was highlighted by the prototype algorithm was that of ``chaos"; ants would frequently deposit an object and then immediately pick it up again, as they were still in contact with it. Conversely, ants would also frequently deposit objects after taking a single ``step", particularly if the placement score was moderately good. This behaviour creates  clusters in which overall placement was quite poor, as large objects were rarely removed from the centre of the cluster (Figure ~\ref{fig:poor}). In order to deal with this problem, a ``cooling down" delay period was introduced; agents must wait a set number of steps after picking up or depositing an object before they carrying out any further placement calculations. This appeared to solve the problem; different (non-zero) values for the delay variable impacted only on the run-time of the simulation, and did not affect the quality of clusters generated. A delay value of 4 was chosen for what follows.

\subsection{The algorithm}

We now describe in detail our algorithm for annular sorting. Ants are modelled by {\it agents}, each of which has the following attributes:

\begin{itemize}
\item Location (application representation)
\item Laden (true/false)
\item Object (application representation of object carried if Laden == true)
\item Energy (integer)
\item Delay (integer)
\end{itemize}

\noindent
Objects are modelled as spheres, and have a single attribute, from which their minimum and maximum perimeters are calculated: 

\begin{itemize}
\item Size (real from the set $[0.5, 1.5, 3]$, corresponding to small, medium or large)
\end{itemize}

\noindent
The following variables are defined to deal with objects (all distances are measured from the edge of objects, rather than from their centre):

\begin{itemize}
\item BASESCORE (real)
\item BONUS (real)
\item PENALTY (integer)
\item BONUSMULTIPLIER (real) (to calculate maximum perimeter)
\item PENALTYMULTIPLIER (real) (to calculate minimum perimeter)
\end{itemize}

There are $n$ objects and $m$ ants initially distributed at random on a two-dimensional ``board" of fixed size. The number of ants and numbers of objects of each size may be specified in advance. Ants may move over other ants or over objects; there are no spatial restrictions other than those imposed by collisions between unladen ants and objects.

\newpage
\noindent
The pseudo-code expression of the algorithm is as follows:

\begin{tabbing}
.......\= .......\= .......\= .......\= .......\= \kill
{\bf while} (agents exist)\\
\\
	\> {\bf for all} agents ($A_1, \dots A_i, \dots A_m$) do \\
\\
		\> \> // Remove ant if ``dead" \\
		\> \> {\bf if} ($A_i$.Energy == 0) \\
		\> \>	\> remove $A_i$ from system \\
		\> \>	\> {\bf break} // i.e. go to next ant, if possible  \\
		\> \> {\bf end if} \\
\\
		\> \> // Check delay\\
		\> \> {\bf if} ($A_i$.Delay $>$ 0)\\
		\> \> 	\> $A_i$.Delay = $A_i$.Delay -1 \\
		\> \> {\bf else}
\\
		\> \> \> // Unladen ant collides with object \\
		\> \> \> {\bf if} ($A_i$.Laden == false {\bf and} ($A_i$.Location == some $O_i$.Location)) \\
		\> \> \>	\> Score = CalculatePlacement($O_i$) \\
		\> \> \>	\> Probability = random$(0 \dots 1$) \\
		\> \> \>	\> {\bf if} (Score $>$ Probability) \\
		\> \> \>	\> 	\> $A_i$.Delay = 4 \\
		\> \> \>	\>	\> $A_i$.Object = $O_i$ \\
		\> \> \>	\>	\> $A_i$.Laden = true \\
		\> \> \>	\>	\> $A_i$.Energy = $A_i$.Energy - 1\\
		\> \> \>	\> {\bf end if} \\
		\> \> \> {\bf end if} \\
\\
		\> \> \> // Laden ant in free space \\
		\> \> \> {\bf if} ($A_i$.Laden == true {\bf and} ($A_i$.Location == empty))\\
		\> \> \>	\> Score = CalculatePlacement($A_i$.Object) \\
		\> \> \>	\> Probability = random$(0 \dots 1$) \\
		\> \> \>	\> {\bf if} (Score $>$ Probability) \\
		\> \> \>	\>	\> Place $A_i$.Object at $A_i$.Location \\
		\> \> \>	\>	\> $A_i$.Delay = 4 \\
		\> \> \>	\>	\> $A_i$.Laden = false \\
		\> \> \>	\> {\bf end if}\\
		\> \> \> {\bf end if}\\

		\> \> {\bf end if}\\
\\
		\> \> Move $A_i$ to randomly selected adjacent location \\
\\
	\> {\bf {\bf end for}} \\
{\bf end while}\\

\end{tabbing}
\newpage

\noindent
We now describe the CalculatePlacement() function:

\begin{tabbing}
.......\= .......\= .......\= .......\= \kill
CalculatePlacement(Object)\\
\\
Score = BASESCORE\\
\\
{\bf for each} neighbouring Object $N_i$\\
\\
	\> // ``within" calculates Cartesian proximity \\
	\> {\bf if} ($N_i$.Location within (PENALTYMULTIPLIER * Object.Size))\\
	\>	\> Score = Score - PENALTY \\
	\> {\bf else} \\
	\> {\bf if} ($N_i$.Location within (BONUSMULTIPLIER * Object.Size))\\
	\>	\> Score = Score + BONUS\\
	\>{\bf end if} \\
\\
{\bf {\bf end for}}\\
\\
{\bf return} (Score)\\
\end{tabbing}
	
\subsection{Implementation}

The pseudo-code above was implemented using the Breve\footnote{Available at http://www.spiderland.org/breve/}
multi-agent system environment \cite{breve}. This package facilitates the simulation of decentralised systems, in a similar fashion to the well-known Swarm \footnote{http://www.swarm.org} system. The benefits of using such a framework are derived from the fact that it automatically handles issues such as collision detection, construction of object neighbourhoods, and the user interface. This allows the programmer to concentrate effort on the important aspects of the model's implementation, rather than on ``house keeping" issues. All simulations were run on an Apple Macintosh under MacOS, although the code runs equally well on both Linux and Windows machines. An additional Java application was written to deal with post-processing of object coordinates. All code is available, on request, from the corresponding author (MA).

\section{Results}

\begin{figure}
\begin{center}
\includegraphics[scale=0.3]{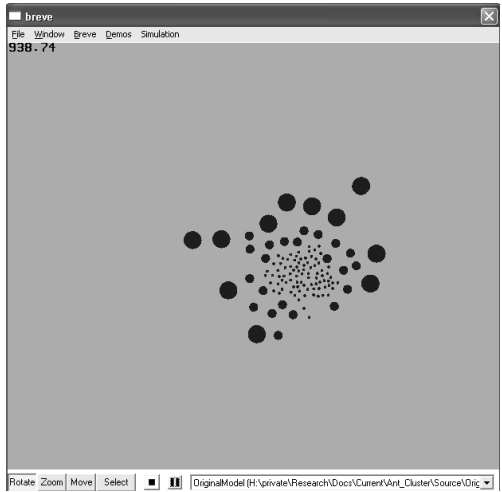}
\caption{Typical final cluster pattern}
\label{fig:typical}
\end{center}
\end{figure}

We found that the simulation consistently generated final cluster patterns that are strongly reminiscent of those observed in real colonies by Franks {\it et al.} \cite{franks02}. A typical final pattern is depicted in Figure ~\ref{fig:typical}; note
the densely-packed core of smaller items at the centre of the cluster, followed by successive ``rings" of increasingly larger and more widely-spaced items, as observed in Figure ~\ref{fig:franks}.

In order to compare our simulation results with those of Franks {\it et al.} \cite{franks02}, we carried out two experiments. In Experiment 1 we ran the simulation 50 times with the following parameter values:

\begin{itemize}
\item Number of small brood items = 25
\item Number of medium brood items = 25
\item Number of large brood items = 25
\item BONUSMULTIPLIER = 4.0
\item PENALTYMULTIPLIER = 0.4
\item Base score = 0
\item PENALTY = -50
\item BONUS = 0.1
\end{itemize}

In Experiment 2, we carried out an additional 50 runs, replacing the following parameter values to reflect the object distribution of Colony A in \cite{franks02}:

\begin{itemize}
\item Number of small brood items = 85
\item Number of medium brood items = 22
\item Number of large brood items = 11
\end{itemize}

For each experiment, we calculated the mean distance from centroid for each object type. Direct comparisons with the biological study were not possible, due to the different spatial units used; however, dividing the results for distance from centroid by the value of the smallest object gives the ratios described in Table ~\ref{table:results}. Values for Franks 1 refer to Colony A before migration, and Franks 2 refers to Colony A {\it after} migration.

\begin{table}
\begin{center}
\begin{tabular}{|l|l|l|l|l|} \hline
Size & Franks 1 & Franks 2 & Exp. 1 & Exp. 2 \\
\hline
Small     & 1 & 1 & 1 & 1 \\ \hline
Medium & 1.429 & 1.668 & 1.371 & 1.760 \\ \hline
Large    & 2.679 & 2.047 & 2.391 & 2.533 \\ 
\hline
\end{tabular}
\caption{Summary of results}
\label{table:results}
\end{center}
\end{table}

\subsection{Discussion}

Comparing the values, we see that the model achieves values for distance from centroid that are very close to those observed in the biological study. The difference between the simulation and reality is about the same as the difference due to migration in the two biological studies.

We also carried out tile size analysis of our results, in order to assess the distribution of objects of different types. However, the only Voronoi tesselation software available to us (VpPlants\footnote{http://pages.cpsc.ucalgary.ca/~marina/vpplants/}) did not allow the imposition of a bounding box, so we were unable to obtain meaningful comparisons with the tiling results described by Franks {\it et al.} \cite{franks02}.

This study has demonstrated a minimalistic model that can consistently sort 
objects into annular clusters. There are real world examples where this 
might be useful, particularly when one has objects that may change over time 
and require different amounts of separation. Some of the aspects that set 
this study apart from others such as Wilson \textit{et al.} \cite{wilson04} also make it harder to 
create a physical implementation. In particular the model relies on the fact 
that agents are able to pull poorly placed objects out of the centre of a 
cluster without disturbing other objects or risking collision, something 
that is hard to do with simple robotics. These difficulties will become less 
significant as the state of the art advances.

The model proposed in this study works well as an explanation of the sorting 
behaviour observed in \textit{Leptothorax} ants. Unfortunately it is not possible to compare the 
results with those given by Wilson \textit{et al.} \cite{wilson04} directly as a different performance 
metric was used. However the model described in this paper does not make use 
of a complex ``leaky integrator'' system, or have the unrealistic 
constraints on movement that were introduced by the robotic implementation 
used by Wilson \textit{et al.} \cite{wilson04}. 

Towards the end of this study an additional paper by Sendova-Franks {\it et al.} \cite{sendova04}
became available. It presented a further biological study of brood sorting 
by \textit{Leptothorax} Ants. As with the previous study \cite{franks02}, ants were induced into nest 
migration and observed. This time the focus was on the behaviour of the ants 
and the authors were able to observe two distinct phases. During 
the ``clustering'' phase, ants carried brood from the old hive to the new 
one. During this phase carrying distance was determined by brood size. Ants 
formed a small cluster inside the new hive with small objects furthest from 
the entrance and large objects closest. In the second phase the ants spread 
out the pre-clustered brood items to achieve an annular cluster. This work 
naturally has some impact on the model proposed in 
this study but it does not necessarily invalidate it. The behaviour 
described in this paper could well explain what occurs during the clustering 
phase seen in \cite{sendova04}, the only difference being that the ants 
start the second phase with the brood already partially clustered. At present it is not clear 
whether the first phase is actually essential to the process.

\section{Conclusions}

In this paper we have described a minimal model of annular brood sorting in {\it Leptothorax} ants. Simulation results are consistent with the structures observed in actual ant colonies. 
Our model uses only stochastic ant behaviour and a brood item ``repulsion" and ``attraction" mechanism, which supports the previous argument that this emergent behaviour may well be linked to pheromone deposition. Further work in this area will include a more detailed analysis of the results in terms of object distribution, as well as continued experimental investigations into the underlying biological mechanisms.
\bibliography{ants}

\end{document}